\documentclass[twocolumn]{aastex631}
\usepackage{multirow}
\usepackage{float}
\usepackage{subfigure}
\usepackage{booktabs,xltabular}
\usepackage{tabularx}
\usepackage{dblfloatfix}
\usepackage{supertabular}
\usepackage{graphicx}

{\setlength\tabcolsep{3.5pt}}

\begin{document}

\title{The Atacama Cosmology Telescope: Systematic Transient Search of 3-Day Maps}

\author[0000-0001-8093-2534]{Yaqiong~Li}
\affiliation{Department of Physics, Cornell University, Ithaca, NY 14853, USA}
\affiliation{Kavli Institute at Cornell for Nanoscale Science, Cornell University, Ithaca, NY 14853, USA}

\author[0000-0002-2840-9794]{Emily~Biermann}
\affiliation{Department of Physics and Astronomy, University of Pittsburgh, Pittsburgh, PA, 15213, USA}

\author[0000-0002-4478-7111]{Sigurd~Naess}
\affiliation{Institute of theoretical astrophycis, University of Oslo, Norway}

\author[0000-0002-1035-1854]{Simone~Aiola}
\affiliation{Center for Computational Astrophysics, Flatiron Institute, 162 5th Avenue, New York, NY 10010 USA}

\author{Rui~An}
\affiliation{Department of Physics and Astronomy, University of Southern California, Los Angeles, CA 90089, USA}

\author[0000-0001-5846-0411]{Nicholas~Battaglia}
\affiliation{Department of Astronomy, Cornell University, Ithaca, NY 14853, USA}

\author[0000-0002-2971-1776]{Tanay~Bhandarkar}
\affiliation{Department of Physics and Astronomy, University of Pennsylvania, 209 S. 33rd Street, Philadelphia, PA 19104, USA}

\author[0000-0003-0837-0068]{Erminia~Calabrese}
\affiliation{School of Physics and Astronomy, Cardiff University, The Parade, Cardiff, Wales CF24 3AA, UK}

\author[0000-0002-9113-7058]{Steve~K.~Choi}
\affiliation{Department of Physics, Cornell University, Ithaca, NY 14853, USA}
\affiliation{Department of Astronomy, Cornell University, Ithaca, NY 14853, USA}

\author[0000-0001-5068-1295]{Kevin~T.~Crowley}
\affiliation{Center for Astrophysics and Space Studies; University of California, San Diego, SERF Building Room 333; 9500 Gilman Dr, La Jolla, CA 92093-4204, USA}

\author[0000-0002-3169-9761]{Mark~Devlin}
\affiliation{Department of Physics and Astronomy, University of Pennsylvania, 209 S. 33rd Street, Philadelphia, PA 19104, USA}

\author[0000-0002-6318-1924]{Cody~J.~Duell}
\affiliation{Department of Physics, Cornell University, Ithaca, NY 14853, USA}

\author[0000-0002-9693-4478]{Shannon~M.~Duff}
\affiliation{National Institute of Standards and Technology, Quantum Sensors Group, 325 Broadway, Boulder, CO 80305}

\author[0000-0002-7450-2586]{Jo~Dunkley}
\affiliation{Joseph Henry Laboratories of Physics, Jadwin Hall, Princeton University, Princeton, NJ 08544, USA}
\affiliation{Department of Astrophysical Sciences, Peyton Hall,  Princeton University, Princeton, NJ 08544 USA}

\author[0000-0003-3892-1860]{Rolando~Dünner}
\affiliation{Instituto de Astrof\'isica and Centro de Astro-Ingenier\'ia, Facultad de F\'isica, Pontificia Universidad Cat\'olica de Chile, Av. Vicu\~na Mackenna 4860, 7820436 Macul, Santiago, Chile}

\author[0000-0001-9731-3617]{Patricio~A.~Gallardo}
\affiliation{Kavli Institute for Cosmological Physics, University of Chicago, 5640 S Ellis Ave., Chicago, IL 60637, USA}

\author[0000-0002-1697-3080]{Yilun~Guan}
\affiliation{David A. Dunlap Department of Astronomy \&Astrophysics, University of Toronto, 50 St. George St., Toronto ON M5S 3H4, Canada}

\author[0000-0002-4765-3426]{Carlos~Herv\'ias-Caimapo}
\affiliation{Instituto de Astrof\'isica and Centro de Astro-Ingenier\'ia, Facultad de F\'isica, Pontificia Universidad Cat\'olica de Chile, Av. Vicu\~na Mackenna 4860, 7820436 Macul, Santiago, Chile}

\author[0000-0003-1690-6678]{Adam~D.~Hincks}
\affiliation{David A. Dunlap Department of Astronomy \&Astrophysics, University of Toronto, 50 St. George St., Toronto ON M5S 3H4, Canada}
\affiliation{Specola Vaticana (Vatican Observatory), V-00120 Vatican City State }

\author[0000-0002-2781-9302]{Johannes~Hubmayr}
\affiliation{National Institute of Standards and Technology, Quantum Sensors Group, 325 Broadway, Boulder, CO 80305}

\author[0000-0001-7109-0099]{Kevin M. Huffenberger}
\affiliation{Department of Physics, Florida State University, Tallahassee, FL 32306, USA}

\author[0000-0002-8816-6800]{John~P.~Hughes}
\affiliation{Department of Physics and Astronomy, Rutgers University, 136 Frelinghuysen Road, Piscataway, NJ 08854, USA}

\author[0000-0002-3734-331X]{Arthur~Kosowsky}
\affiliation{Department of Physics and Astronomy, University of Pittsburgh, Pittsburgh, PA, 15213, USA}

\author[0000-0002-6849-4217]{Thibaut~Louis}
\affiliation{Université Paris-Saclay, CNRS/IN2P3, IJCLab, 91405 Orsay, France}

\author[0000-0002-2018-3807]{Maya~Mallaby-Kay}
\affiliation{Department of Astronomy and Astrophysics, University of Chicago, 5640 S Ellis Ave, Chicago, IL 60637, USA}

\author[0000-0002-7245-4541]{Jeff~McMahon}
\affiliation{Department of Astronomy and Astrophysics, University of Chicago, 5640 S Ellis Ave, Chicago, IL 60637, USA}
\affiliation{Department of Physics, University of Chicago, 5720 S Ellis Ave, Chicago, IL 60637, USA}
\affiliation{Kavli Institute for Cosmological Physics, University of Chicago, 5640 S Ellis Ave., Chicago, IL 60637, USA}
\affiliation{Enrico Fermi Institute, University of Chicago, 5640 S Ellis Ave, Chicago, IL 60637, USA}
\affiliation{Fermi National Accelerator Laboratory, Batavia, IL 60637, USA}

\author[0000-0002-8307-5088]{Federico~Nati}
\affiliation{University of Milano-Bicocca}

\author[0000-0001-7125-3580]{Michael~D.~Niemack}
\affiliation{Department of Physics, Cornell University, Ithaca, NY 14853, USA}
\affiliation{Department of Astronomy, Cornell University, Ithaca, NY 14853, USA}

\author[0000-0003-1842-8104]{John~Orlowski-Scherer}
\affiliation{Department of Physics, McGill University, 3600 University Street, Montreal, QC, H3A 2T8, Canada}

\author[0000-0002-9828-3525]{Lyman~Page}
\affil{Joseph Henry Laboratories of Physics, Jadwin Hall, Princeton University, Princeton, NJ 08544, USA}

\author[0000-0003-4006-1134]{Maria~Salatino}
\affiliation{Physics Department Stanford University 94305 Stanford CA, US}
\affiliation{Kavli Institute for Astroparticle Physics and Cosmology (KIPAC) 94305 Stanford CA, US}

\author[0000-0002-8149-1352]{Crist\'obal~Sif\'on}
\affiliation{Instituto de F\'isica, Pontificia Universidad Cat\'olica de Valpara\'iso, Casilla 4059, Valpara\'iso, Chile}

\author[0000-0002-7020-7301]{Suzanne T. Staggs}
\affiliation{Joseph Henry Laboratories of Physics, Jadwin Hall, Princeton University, Princeton, NJ 08544, USA}

\author[0000-0001-5327-1400]{Cristian~Vargas}
\affiliation{Instituto de Astrof\'isica and Centro de Astro-Ingenier\'ia, Facultad de F\'isica, Pontificia Universidad Cat\'olica de Chile, Av. Vicu\~na Mackenna 4860, 7820436 Macul, Santiago, Chile}

\author[0000-0002-2105-7589]{Eve~M.~Vavagiakis}
\affiliation{Department of Physics, Cornell University, Ithaca, NY 14853, USA}

\author[0000-0002-8710-0914]{Yuhan~Wang}
\affiliation{Joseph Henry Laboratories of Physics, Jadwin Hall, Princeton University, Princeton, NJ 08544, USA}

\author[0000-0002-7567-4451]{Edward~J.~Wollack}
\affiliation{NASA Goddard Space Flight Center, 8800 Greenbelt Rd, Greenbelt, MD 20771, USA}



\begin{abstract}

We conduct a systematic search for transients in three years of data (2017--2019) from the Atacama Cosmology Telescope (ACT). ACT covers 40 percent of the sky at three bands spanning from 77~GHz to 277~GHz. Analysis of 3-day mean-subtracted sky maps, which were match-filtered for point sources, yielded 29 transients detections. Eight of these transients are due to known asteroids, and three others were previously published. Four of these events occur in areas of with poor noise models and thus we cannot be confident they are real transients. We are left with 14 new transient events occurring at 11 unique locations. All of these events are associated with either rotationally variable stars or cool stars. Ten events have flat or falling spectra indicating radiation from synchrotron emission. One event has a rising spectrum indicating a different engine for the flare. 

\end{abstract}




\section{Introduction} \label{sec:intro}
Transient astronomical sources comprise some of the most dramatic and extreme astrophysical events, such as supernovae, binary star mergers, active galactic nuclei (AGN), gamma ray bursts (GRB), and tidal distruption events (TDE). Continuous monitoring of large sky regions to detect transients has been a major motivation behind the upcoming Rubin Observatory \citep{Rubin} and several other efforts at optical wavelengths, such as the Zwicky Transient Facility \citep{Masci_2018} and ASAS-SN \citep{ASAS-SN}. Transient astronomy has also received a large increase in attention with the first gravitational wave sources detected by LIGO \citep{LIGO_first}, and detection of sources in both gravitational waves and multiple electromagnetic wave bands. Wide-field millimeter wavelength surveys have focused on mapping the unchanging microwave background for cosmological applications, but recently some have reached the combination of high angular resolution, wide survey area and sensitivity required to start probing the time-variable millimeter sky.


In recent years there have been a growing number of millimeter transient detections.  In 2016, the South Pole Telescope (SPT) reported one candidate event observed during the 2012--2013 season at a $2.6\sigma$ significance with a peak flux of $16.5\pm2.4$ mJy in the 150 GHz band~\citep{Whitehorn_2016}. This event was broadly consistent with a GRB orphan afterglow. \cite{Kuno_2004} reported a detection at 90 GHz of the afterglow of GRB 030329 ($z=0.17$) with a flux around 65mJ. More than a decade later, \cite{Laskar_2019} detected a polarized reverse shock from GRB 190114C at a frequency of 97.5 GHz. Other types types of events are also seen in the millimeter. For instance, \cite{Yuan_2016} reported millimeter emission from the TDE IGR J12580+0134. \cite{Ho_2019} studied the unprecedented transient AT2018cow in radio and submillimeter bands, finding its emission to be inconsistent with synchrotron emission or Compton scattering of UVIOR photons, indicating a new class of millimeter transients.

There are also many examples of stellar flares seen in the millimeter. ACT reported three bright transient sources coincident with stars \citep{Naess_2021a} and SPT reported 15 additional transient events~\citep{Guns_2021}. Most of the sources reported from SPT were consistent with stellar activity, except for two of their events which may be extragalactic in origin. Millimeter stellar flares have also been reported in other works such as \cite{Bower_2003}, \cite{Brown_2006}, \cite{Massi_2006}, \cite{Mairs_2019} and are most likely due to stellar magnetic field reconnection.

In this work, we complete a systematic search for transient events with three years' of ACT data using 3-day stacked maps. In Section~\ref{sec:data}, we introduce the ACT survey and the maps used for this analysis. In Section~\ref{sec:method}, we describe how the transient events are detected and how data cuts applied to the sample. In Section~\ref{sec:results}, we present our findings and possible counterparts for each transient. After all cuts are applied we are left with 29 events. Three of these events are redetections from \cite{Naess_2021a}. Eight appear to be from asteroids that were not masked in the 3-day maps and four occur in areas with poor noise models and cannot be confirmed as real transients. This leaves 14 new transient detections at 11 unique positions on the sky. Most of the events have clear stellar associations. In Section~\ref{sec:concl}, we summarize the nature of these counterparts.

\section{Data} \label{sec:data}

The Atacama Cosmology Telescope (ACT) was a 6~m off-axis Gregorian telescope located in the Atacama Desert in Chile at an altitude of 5200~m \citep{Fowler:07}. AdvACT was the third generation ACT instrument.  It housed three optics tubes with a total field of view (FOV) spanning ${\sim}1.5^{\circ}$\citep{Thornton_2016}, each containing a set of lenses, low-pass filters, and a single AdvACT detector array \citep{Ho_2017,Choi_2018}. The ACT dataset considered here covers the three bandpasses f090 (77--112 GHz), f150 (124--172 GHz) and f220 (182--277 GHz) using the three dichroic polarization-sensitive transition edge sensor (TES) bolometer arrays PA4 (f150+f220), PA5 (f090+f150) and PA6 (f090+f150).
During observations, the telescope scanned the sky in azimuth at a fixed elevation at a scan speed of ${\sim}1.5$~deg per second. When the sky was rising, it took ${\sim}$6 minutes for a point on the sky to be gradually swept across by the detectors in PA4 and PA5. Then, after $\sim$3 minutes this process would repeat for the detectors in PA6. This order was reversed when the sky was setting. The bolometer signals were stored in time-ordered data (TOD) files, each containing a roughly 10 minute time series for all detectors in a single array. Detailed data selection and calibration process is described in \cite{Aiola_2020}.

We search for transient events using 3-day maps originally created to search for Planet 9 \citep{Naess_2021b}. The search covered 18000\,deg$^{2}$ of the sky using ACT data taken from 2017 to 2019. Although these maps were not designed for a systematic transient search, their short timescale is well suited for this purpose. The maps are made by dividing the daytime and nighttime TODs taken by each of the three detector array at each of the two frequency bands into approximately 3-day long data chunks. Therefore, there are twelve maps (day and night time data within three arrays with two bands) for each 3-day period. The maps are made by calculating the maximum likelihood estimate for the CMB temperature from TOD. The map is then matched filtered to look for point sources, which gives an estimation of flux density and $S/N$ at each pixel. Detailed methods are described in \cite{Naess_2021b}.  For each 3-day map, a mean sky subtraction at the corresponding frequency band is implemented so only the time-varying signal remains. Regions around the planets and bright asteroids are masked to avoid contamination from planet sidelobes and false transient detection from known moving astronomical objects. 
Since the 3-day maps were made as part of a search for Planet 9, some tradeoffs are made that are suitable when performing large stacks of maps, but at times not ideal when analyzing maps in isolation. In particular the maps' matched filter numerator was computed directly in time domain while the denominator was computed in map-space. This approach is fast, but only accurate if the time-domain noise model and map-space noise model are consistent. In practice this is an approximation which introduces a bias in areas where the map hit-density changes rapidly from pixel to pixel. This does not happen when performing large stacks like in the Planet 9 search, but it is quite common when looking for objects in the individual 3-day maps, resulting in many spurious detections (see Section~\ref{subsec:cuts}). Furthermore, the matched filtered 3-day maps are stored at 1 arcminute resolution, which was an acceptable performance tradeoff for the Planet 9 search (where this was a minor contribution to the overall smoothing budget), but is suboptimal for blind transient detection. The ways in which the position uncertainty introduced by the selected resolution are handled is detailed in Section~\ref{sub:characterization}. We plan to resolve these deficiencies in a future paper using maps tailored for transient detection.


\section{Methodology} \label{sec:method}
\subsection{Initial Detection and Spurious Candidate Cuts}
\label{subsec:cuts}

\begin{figure}[H]
    \centering
    \includegraphics[width=0.45\textwidth]{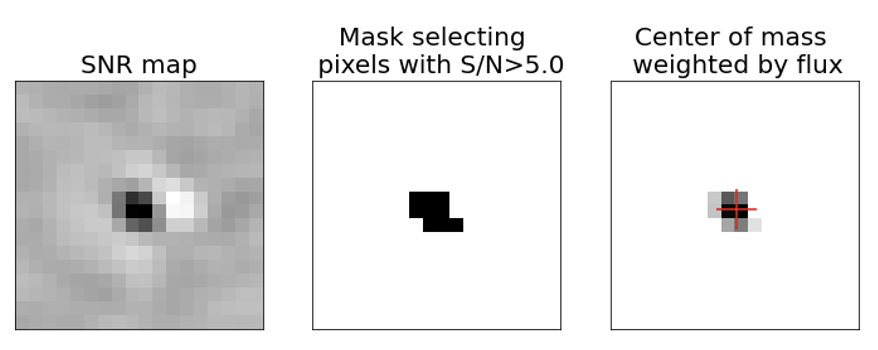}
    \caption{Process of the initial detection, with each plot showing a 0.3 deg by 0.3 deg map. The first step is to make a mask (middle) on $S/N$ map (left) selecting pixels that have $S/N >5$. The mask is then applied to the flux map (right), and the candidate position, shown as the red cross mark, is evaluated as the center of mass weighted by the flux values within the selected pixels. }
    \label{fig:initial_det}
\end{figure}
Any pixel or group of pixels with $S/N >5$ in the 3-day sky-subtracted maps is considered a transient candidate. For each set of 3-day S/N  and flux density maps, we first mask pixels with $S/N<5$, then apply this mask to the corresponding flux density map (Figure~\ref{fig:initial_det}). The positions of candidates are evaluated to be the flux-weighted centroids. This initial detection finds 332,333 candidate events. If a detection is a real transient event and not simply a noise fluctuation, we will likely detect it in more than one array. Thus, we cross match each candidate between detector arrays with a matching distance of 1.5 arcminutes, which is 1.5 times the resolution of the 3-day maps. We only keep candidates that appear in at least two frequency and array combinations. This cuts $76\%$ of the initial detections. 

Even with this cut applied, there are still many spurious detections grouped in clusters and along map edges. We perform three additional geometric cuts, requiring a candidate to be cut in all detected frequency and array combinations to exclude it from the analysis. The results of these cuts are summarized in Table~\ref{tab:cut_stats}.\footnote{Note that a candidate might be cut by more than one criteria} We apply these cuts in tandem to the frequency and array requirement, thus the following statistics include all candidates.

\begin{figure*}[!htbp]
   \centering
   \includegraphics[width=0.3\linewidth]{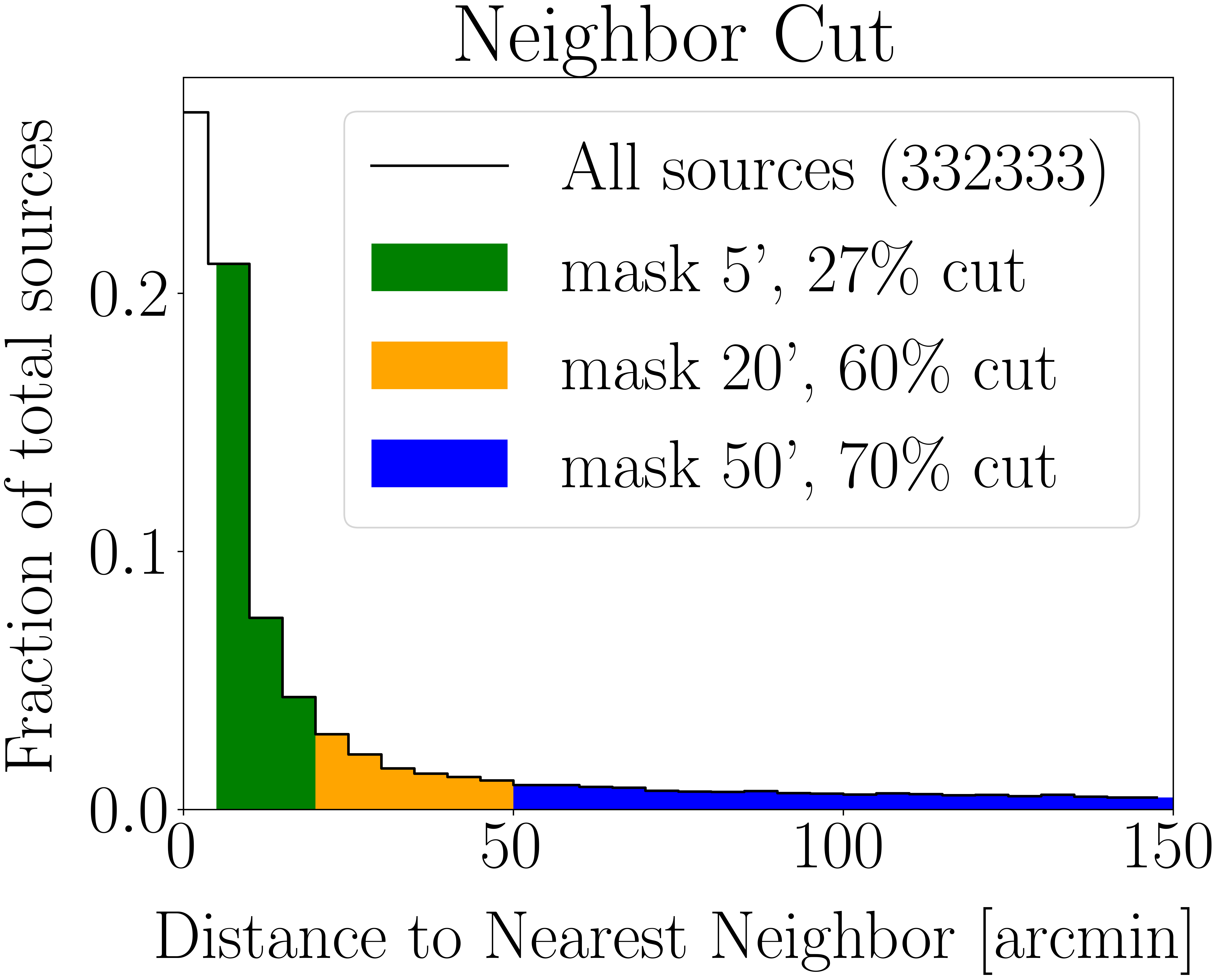}
    \includegraphics[width=0.3\linewidth]{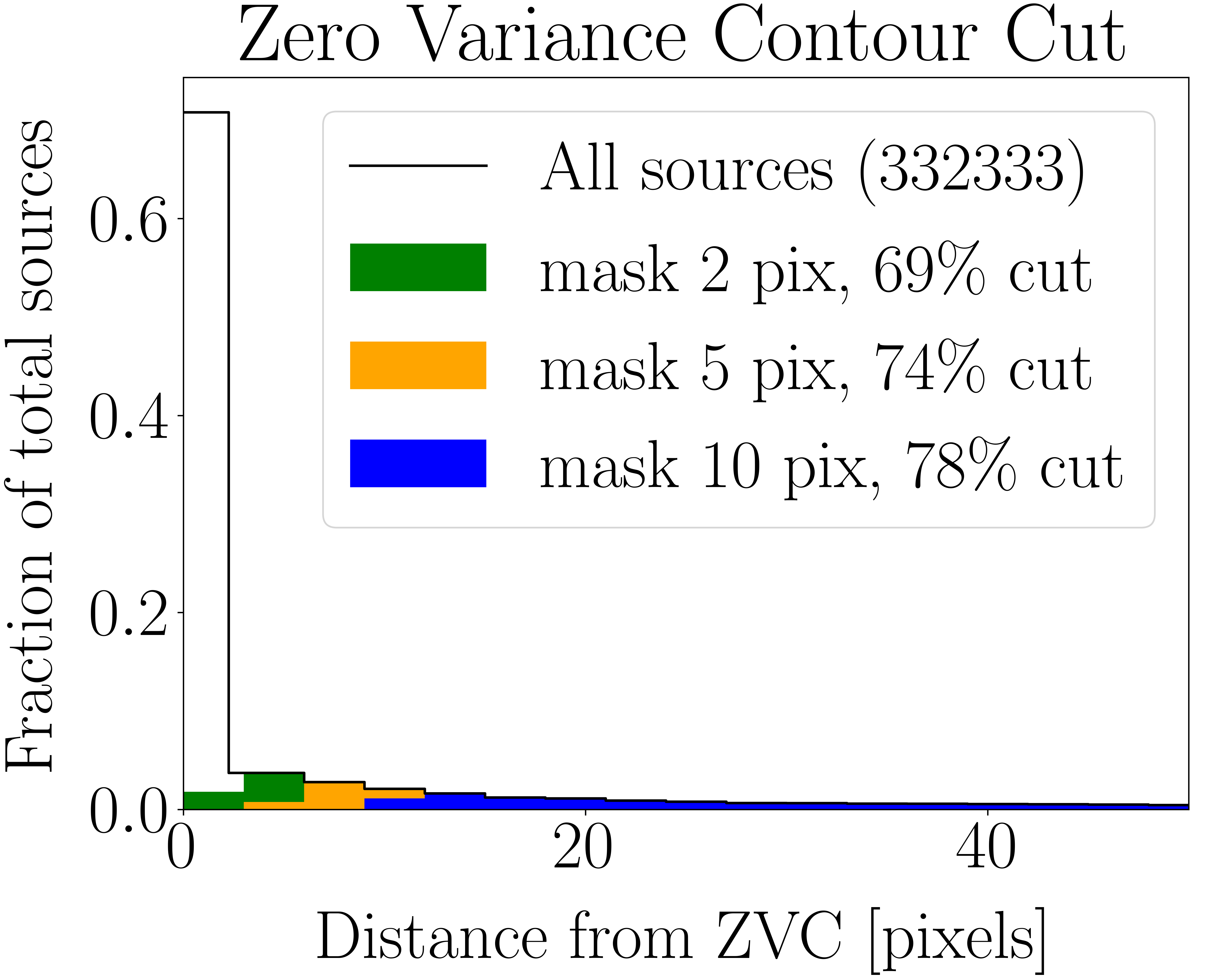}
    \includegraphics[width=0.3\linewidth]{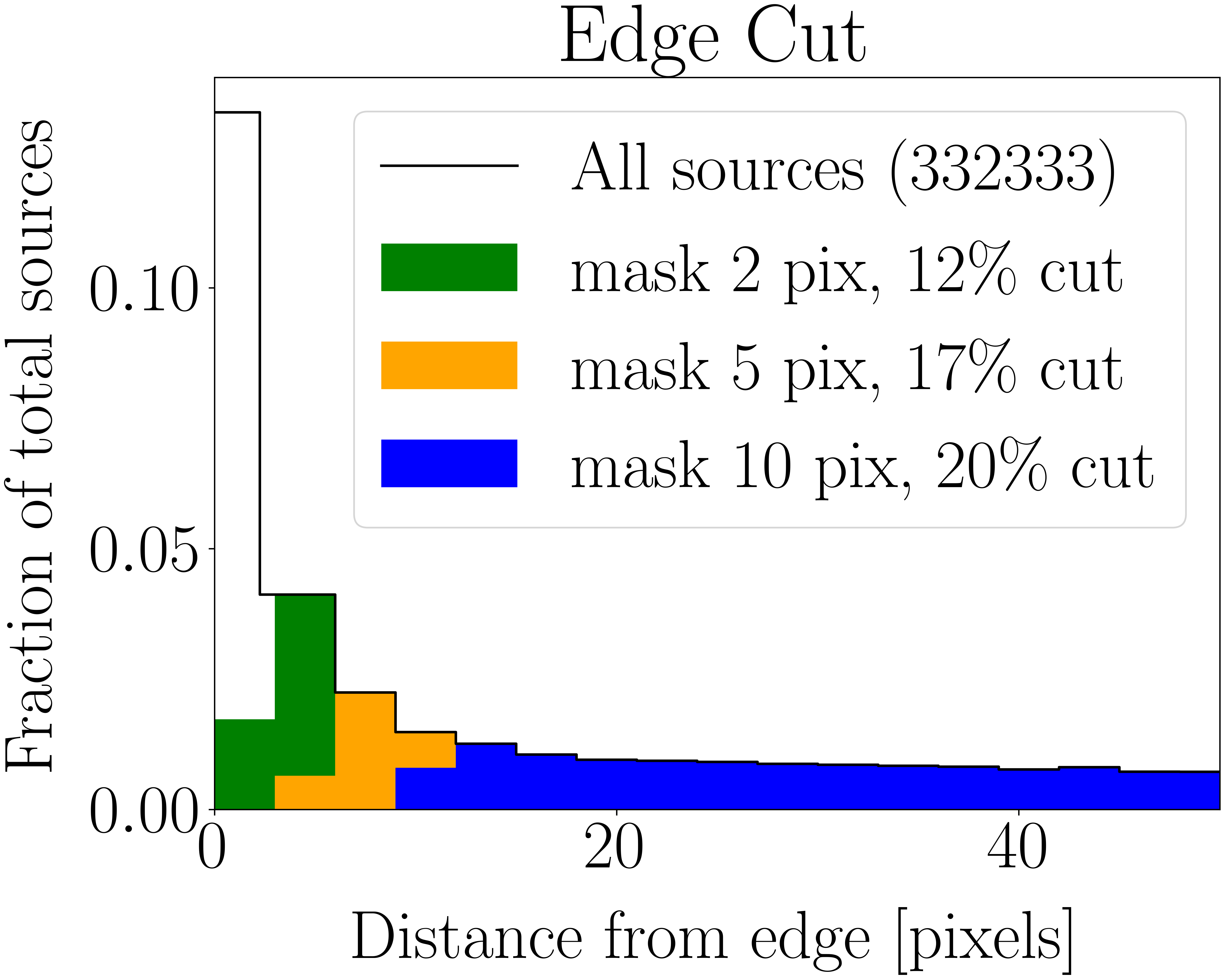}
    \caption{{\bf \textit{Left:}} A histogram of the distance to each detection's nearest neighbor with a binsize of 5 arcminutes. The peak close to zero indicates there are clusters of spurious detections in many of the maps so we cut any candidate with a nearest neighbor of 20 arcminutes or less. {\bf \textit{Center:}} A histogram of each candidate's distance from the nearest zero variance contour, defined to be a variance of less than $1.5\times10^{-5} K$, with a bin size of 3 pixels. There is a large peak of candidates near zero variance contours which drops after 3 pixels. We mask out to 5 pixels, cutting $74\%$ of all sources. {\bf \textit{Right:}} A histogram of each candidate's distance from the map edge in pixel units with a bin size of 3 pixels. As expected, there is an excess number of candidates near the edge of the map as the map edges are noisy and so appear variable when sampled every three days. At a mask size of 5 pixels, we cut off the peak of candidates near the edge.}
    \label{fig:histcuts}
\end{figure*}

The first geometric cut is motivated by the observation that many of the 3-day exhibit a stripy pattern of spurious detections along the scanning direction. These false detections are caused by a systematic underestimation of the variance when the coverage is uneven due to low hit counts. The stripy regions are reliably identified by searching for detections with nearby neighbours within the same 3-day day map. This approach is motivated by the lefthand plot in Figure~\ref{fig:histcuts} which shows a histogram of each candidate's nearest neighbour: there is a clear excess of detections with neighbours within $\sim$1\,deg, which would not be expected for real detections since extragalactic transients should be spatially uncorrelated. Based on this histogram, we cut any candidates with a neighbor within 20 arcminutes as there is an overdensity of detections with a closest neighbor within this range. This cut removes $60\%$ of all candidates. 

Second, we see a similar stripy pattern along zero variance contours. During the map making process, some approximations result in streaks of spurious detections with vanishing variance and therefore a high $S/N$. In the center plot in Figure~\ref{fig:histcuts} we plot a histogram of the candidates' distances from a zero variance region and cut any candidate within 5 pixels of these features (about $74\%$ of all candidates). 

Lastly, we cut any candidate near a map edge. Map edges are especially noisy and appear to be variable when sampled every three days. In the right-hand plot in Figure~\ref{fig:histcuts} we plot a histogram of the pixel distance of each candidate from the edges of the map. We see a spike in candidates within five pixels from the map edge and so we mask this region. This cuts applies to $17\%$ of all candidates. 

After these cuts are applied an internal cross match is performed to find repeating events. Candidates with positions within one arcminute of each other are considered the same object. This leaves us with 667 independent candidates.

\begin{table*}[htb!]
    \centering
    \begin{tabular}{|c|c|c|c|}
    \hline
                    &  Remaining Candidates    &  Fraction Cut    &  Fraction of map masked  \\
    \hline
    \hline
    frequency and array cut & 78,367        &  0.76               & -- \\
    neighbor cut    &  133,119               &  0.60               &  0.005                   \\
    ZVC cut         &  88,936               &  0.74               &  0.030                   \\
    edge cut        &  278,044               &  0.17               &  0.010                   \\
    \hline
    all cuts        &  5,020                &  0.94               &  0.045                   \\
    \hline
    \end{tabular}
    \caption{This table summarizes the three geometric data cuts applied to each 3-day map after the requirement that a candidate appear in at least two frequency or array combinations is applied. The first column quotes the number of candidates remaining after that cut, the second gives the fraction of all candidates that do not pass these cuts, and the last column gives the fraction of the total number of pixels masked by each cut. Note that these cuts are done in tandem and are independent from each other. The neighbor cut removes candidates with neighbors in the same map within 20 arcminutes, the zero variance contour (ZVC) cut masks candidates within five pixels of zero variance contours, and the edge cut masks candidates within five pixels of the edge.}
\label{tab:cut_stats}
\end{table*}

\subsection{Candidate Verification}
\label{subsec:verfication}


For each of the 667 candidates, we calculate the mean flux by applying a matched filter to the mean sky map using data from 2017 to 2021, and mask candidates with mean flux $>$~50~mJy or $<-50$~mJy. The high-mean-flux candidates are likely variable point sources such as AGN or dusty star forming galaxies (DSFG). These high-mean-flux candidates are already detected in our standard point source catalogs, and there are dedicated studies in preparation on the light curves of these candidates. In this paper we will concentrate on candidates that are not detectable in the mean sky maps. The candidates with negative mean fluxes are located close to bright point sources. The flux density around this region is negative due to the effect of the matched filter that acts as high pass filter, and will increase when the flux of the corresponding point source decreases, resulting in a false detection. There are 82 candidates left after applying the cut on mean flux. We then make light curves to confirm these detections, requiring a $S/N>$~3.5 in at least two arrays or in one array at both frequency bands. These light curves are made from forced photometry, which gives a per-detector flux and flux error, and then evaluating a weighted mean flux as the array-wise result at each frequency band for each scan. This analysis further cuts 41 candidates. 

\begin{figure*}[htbp]
\footnotesize
\centering
\begin{tabular}{|l|llllll|l|l|llllll|}
\hline
& pa4f220 & pa4f150 & pa5f150 & pa6f150 & pa5f090 & pa6f090 & & & pa4f220 & pa4f150 & pa5f150 & pa6f150 & pa5f090 & pa6f090 \\
\hline
1 & \includegraphics[width=0.05\textwidth]{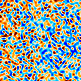} & \includegraphics[width=0.05\textwidth]{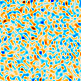} & \includegraphics[width=0.05\textwidth]{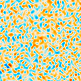} & \includegraphics[width=0.05\textwidth]{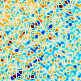} & \includegraphics[width=0.05\textwidth]{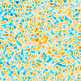} & \includegraphics[width=0.05\textwidth]{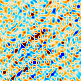} &  &  8 & \includegraphics[width=0.05\textwidth]{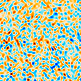} & \includegraphics[width=0.05\textwidth]{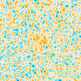} & \includegraphics[width=0.05\textwidth]{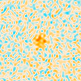} & \includegraphics[width=0.05\textwidth]{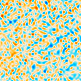} & \includegraphics[width=0.05\textwidth]{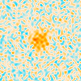} & \includegraphics[width=0.05\textwidth]{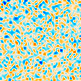} \\
 & \includegraphics[width=0.05\textwidth]{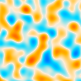} & \includegraphics[width=0.05\textwidth]{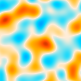} & \includegraphics[width=0.05\textwidth]{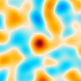} & \includegraphics[width=0.05\textwidth]{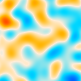} & \includegraphics[width=0.05\textwidth]{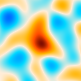} & \includegraphics[width=0.05\textwidth]{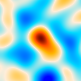} &  &   & \includegraphics[width=0.05\textwidth]{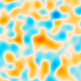} & \includegraphics[width=0.05\textwidth]{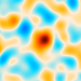} & \includegraphics[width=0.05\textwidth]{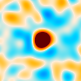} & \includegraphics[width=0.05\textwidth]{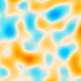} & \includegraphics[width=0.05\textwidth]{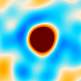} & \includegraphics[width=0.05\textwidth]{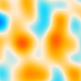} \\
\hline
3 & \includegraphics[width=0.05\textwidth]{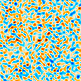} & \includegraphics[width=0.05\textwidth]{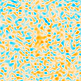} & \includegraphics[width=0.05\textwidth]{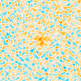} & \includegraphics[width=0.05\textwidth]{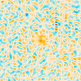} & \includegraphics[width=0.05\textwidth]{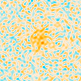} & \includegraphics[width=0.05\textwidth]{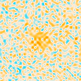} &  &  7 & \includegraphics[width=0.05\textwidth]{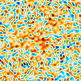} & \includegraphics[width=0.05\textwidth]{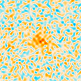} & \includegraphics[width=0.05\textwidth]{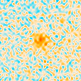} & \includegraphics[width=0.05\textwidth]{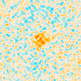} & \includegraphics[width=0.05\textwidth]{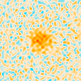} & \includegraphics[width=0.05\textwidth]{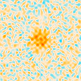} \\
& \includegraphics[width=0.05\textwidth]{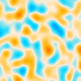} & \includegraphics[width=0.05\textwidth]{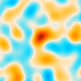} & \includegraphics[width=0.05\textwidth]{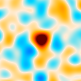} & \includegraphics[width=0.05\textwidth]{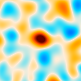} & \includegraphics[width=0.05\textwidth]{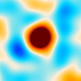} & \includegraphics[width=0.05\textwidth]{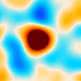} &  &   & \includegraphics[width=0.05\textwidth]{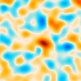} & \includegraphics[width=0.05\textwidth]{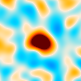} & \includegraphics[width=0.05\textwidth]{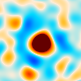} & \includegraphics[width=0.05\textwidth]{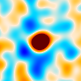} & \includegraphics[width=0.05\textwidth]{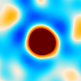} & \includegraphics[width=0.05\textwidth]{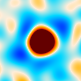} \\
\hline
4a & \includegraphics[width=0.05\textwidth]{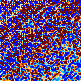} & \includegraphics[width=0.05\textwidth]{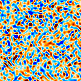} & \includegraphics[width=0.05\textwidth]{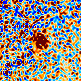} & \includegraphics[width=0.05\textwidth]{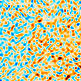} & \includegraphics[width=0.05\textwidth]{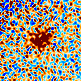} & \includegraphics[width=0.05\textwidth]{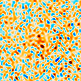} &  &  11 & \includegraphics[width=0.05\textwidth]{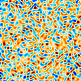} & \includegraphics[width=0.05\textwidth]{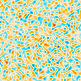} & \includegraphics[width=0.05\textwidth]{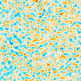} & \includegraphics[width=0.05\textwidth]{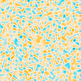} & \includegraphics[width=0.05\textwidth]{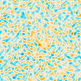} & \includegraphics[width=0.05\textwidth]{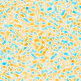} \\
& \includegraphics[width=0.05\textwidth]{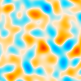} & \includegraphics[width=0.05\textwidth]{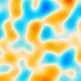} & \includegraphics[width=0.05\textwidth]{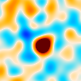} & \includegraphics[width=0.05\textwidth]{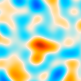} & \includegraphics[width=0.05\textwidth]{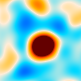} & \includegraphics[width=0.05\textwidth]{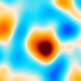} &  &   & \includegraphics[width=0.05\textwidth]{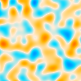} & \includegraphics[width=0.05\textwidth]{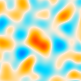} & \includegraphics[width=0.05\textwidth]{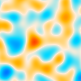} & \includegraphics[width=0.05\textwidth]{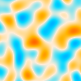} & \includegraphics[width=0.05\textwidth]{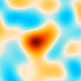} & \includegraphics[width=0.05\textwidth]{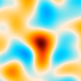} \\
\hline
4b & \includegraphics[width=0.05\textwidth]{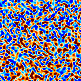} & \includegraphics[width=0.05\textwidth]{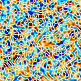} & \includegraphics[width=0.05\textwidth]{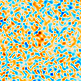} & \includegraphics[width=0.05\textwidth]{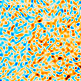} & \includegraphics[width=0.05\textwidth]{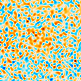} & \includegraphics[width=0.05\textwidth]{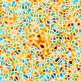} &  &  12a & \includegraphics[width=0.05\textwidth]{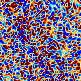} & \includegraphics[width=0.05\textwidth]{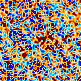} & \includegraphics[width=0.05\textwidth]{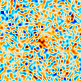} & \includegraphics[width=0.05\textwidth]{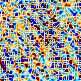} & \includegraphics[width=0.05\textwidth]{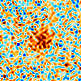} & \includegraphics[width=0.05\textwidth]{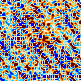} \\
& \includegraphics[width=0.05\textwidth]{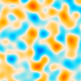} & \includegraphics[width=0.05\textwidth]{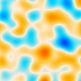} & \includegraphics[width=0.05\textwidth]{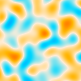} & \includegraphics[width=0.05\textwidth]{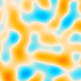} & \includegraphics[width=0.05\textwidth]{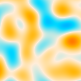} & \includegraphics[width=0.05\textwidth]{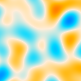} &  &   & \includegraphics[width=0.05\textwidth]{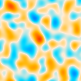} & \includegraphics[width=0.05\textwidth]{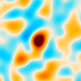} & \includegraphics[width=0.05\textwidth]{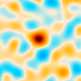} & \includegraphics[width=0.05\textwidth]{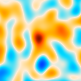} & \includegraphics[width=0.05\textwidth]{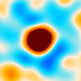} & \includegraphics[width=0.05\textwidth]{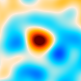} \\
\hline
5a & \includegraphics[width=0.05\textwidth]{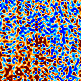} & \includegraphics[width=0.05\textwidth]{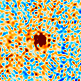} & \includegraphics[width=0.05\textwidth]{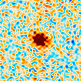} & \includegraphics[width=0.05\textwidth]{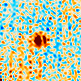} & \includegraphics[width=0.05\textwidth]{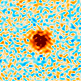} & \includegraphics[width=0.05\textwidth]{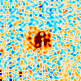} &  &  12b & \includegraphics[width=0.05\textwidth]{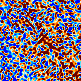} & \includegraphics[width=0.05\textwidth]{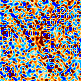} & \includegraphics[width=0.05\textwidth]{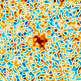} & \includegraphics[width=0.05\textwidth]{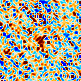} & \includegraphics[width=0.05\textwidth]{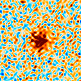} & \includegraphics[width=0.05\textwidth]{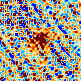} \\
& \includegraphics[width=0.05\textwidth]{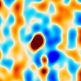} & \includegraphics[width=0.05\textwidth]{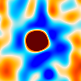} & \includegraphics[width=0.05\textwidth]{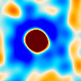} & \includegraphics[width=0.05\textwidth]{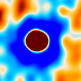} & \includegraphics[width=0.05\textwidth]{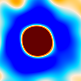} & \includegraphics[width=0.05\textwidth]{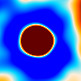} &  &   & \includegraphics[width=0.05\textwidth]{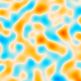} & \includegraphics[width=0.05\textwidth]{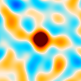} & \includegraphics[width=0.05\textwidth]{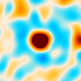} & \includegraphics[width=0.05\textwidth]{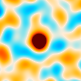} & \includegraphics[width=0.05\textwidth]{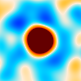} & \includegraphics[width=0.05\textwidth]{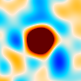} \\
\hline
5b & \includegraphics[width=0.05\textwidth]{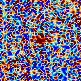} & \includegraphics[width=0.05\textwidth]{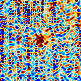} & \includegraphics[width=0.05\textwidth]{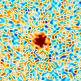} & \includegraphics[width=0.05\textwidth]{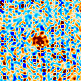} & \includegraphics[width=0.05\textwidth]{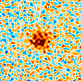} & \includegraphics[width=0.05\textwidth]{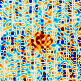} &  &  14 & \includegraphics[width=0.05\textwidth]{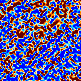} & \includegraphics[width=0.05\textwidth]{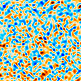} & \includegraphics[width=0.05\textwidth]{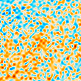} & \includegraphics[width=0.05\textwidth]{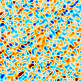} & \includegraphics[width=0.05\textwidth]{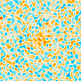} & \includegraphics[width=0.05\textwidth]{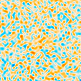} \\
& \includegraphics[width=0.05\textwidth]{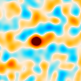} & \includegraphics[width=0.05\textwidth]{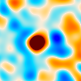} & \includegraphics[width=0.05\textwidth]{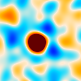} & \includegraphics[width=0.05\textwidth]{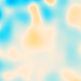} & \includegraphics[width=0.05\textwidth]{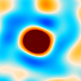} & \includegraphics[width=0.05\textwidth]{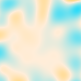} &  &   & \includegraphics[width=0.05\textwidth]{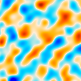} & \includegraphics[width=0.05\textwidth]{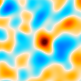} & \includegraphics[width=0.05\textwidth]{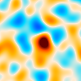} & \includegraphics[width=0.05\textwidth]{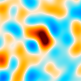} & \includegraphics[width=0.05\textwidth]{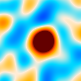} & \includegraphics[width=0.05\textwidth]{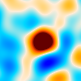} \\
\hline
6 & \includegraphics[width=0.05\textwidth]{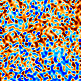} & \includegraphics[width=0.05\textwidth]{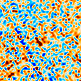} & \includegraphics[width=0.05\textwidth]{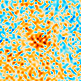} & \includegraphics[width=0.05\textwidth]{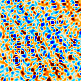} & \includegraphics[width=0.05\textwidth]{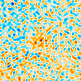} & \includegraphics[width=0.05\textwidth]{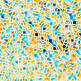} &  &  15 &  & \includegraphics[width=0.05\textwidth]{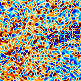} & \includegraphics[width=0.05\textwidth]{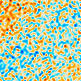} & \includegraphics[width=0.05\textwidth]{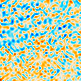} & \includegraphics[width=0.05\textwidth]{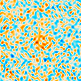} & \includegraphics[width=0.05\textwidth]{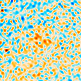} \\
& \includegraphics[width=0.05\textwidth]{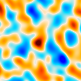} & \includegraphics[width=0.05\textwidth]{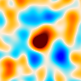} & \includegraphics[width=0.05\textwidth]{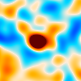} & \includegraphics[width=0.05\textwidth]{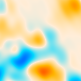} & \includegraphics[width=0.05\textwidth]{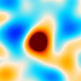} & \includegraphics[width=0.05\textwidth]{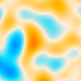} &  &   &  & \includegraphics[width=0.05\textwidth]{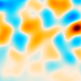} & \includegraphics[width=0.05\textwidth]{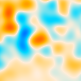} & \includegraphics[width=0.05\textwidth]{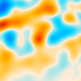} & \includegraphics[width=0.05\textwidth]{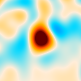} & \includegraphics[width=0.05\textwidth]{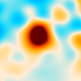} \\
\hline
\hline
2 & \includegraphics[width=0.05\textwidth]{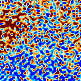} & \includegraphics[width=0.05\textwidth]{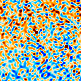} & \includegraphics[width=0.05\textwidth]{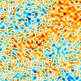} & \includegraphics[width=0.05\textwidth]{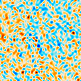} & \includegraphics[width=0.05\textwidth]{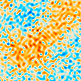} & \includegraphics[width=0.05\textwidth]{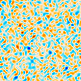} &  &  9 & \includegraphics[width=0.05\textwidth]{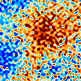} & \includegraphics[width=0.05\textwidth]{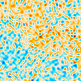} & \includegraphics[width=0.05\textwidth]{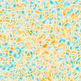} & \includegraphics[width=0.05\textwidth]{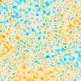} & \includegraphics[width=0.05\textwidth]{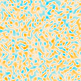} & \includegraphics[width=0.05\textwidth]{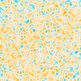} \\
& \includegraphics[width=0.05\textwidth]{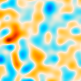} & \includegraphics[width=0.05\textwidth]{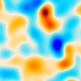} & \includegraphics[width=0.05\textwidth]{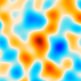} & \includegraphics[width=0.05\textwidth]{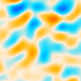} & \includegraphics[width=0.05\textwidth]{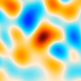} & \includegraphics[width=0.05\textwidth]{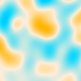} &  &   & \includegraphics[width=0.05\textwidth]{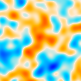} & \includegraphics[width=0.05\textwidth]{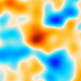} & \includegraphics[width=0.05\textwidth]{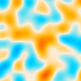} & \includegraphics[width=0.05\textwidth]{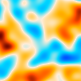} & \includegraphics[width=0.05\textwidth]{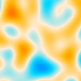} & \includegraphics[width=0.05\textwidth]{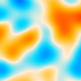} \\
\hline
10 & \includegraphics[width=0.05\textwidth]{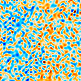} & \includegraphics[width=0.05\textwidth]{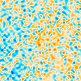} & \includegraphics[width=0.05\textwidth]{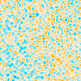} & \includegraphics[width=0.05\textwidth]{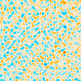} & \includegraphics[width=0.05\textwidth]{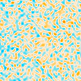} & \includegraphics[width=0.05\textwidth]{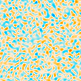} &  &  13 & \includegraphics[width=0.05\textwidth]{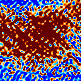} & \includegraphics[width=0.05\textwidth]{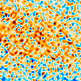} & \includegraphics[width=0.05\textwidth]{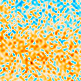} & \includegraphics[width=0.05\textwidth]{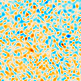} & \includegraphics[width=0.05\textwidth]{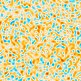} & \includegraphics[width=0.05\textwidth]{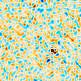} \\
& \includegraphics[width=0.05\textwidth]{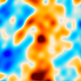} & \includegraphics[width=0.05\textwidth]{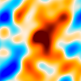} & \includegraphics[width=0.05\textwidth]{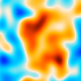} & \includegraphics[width=0.05\textwidth]{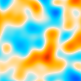} & \includegraphics[width=0.05\textwidth]{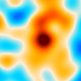} & \includegraphics[width=0.05\textwidth]{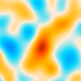} &  &   & \includegraphics[width=0.05\textwidth]{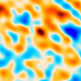} & \includegraphics[width=0.05\textwidth]{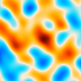} & \includegraphics[width=0.05\textwidth]{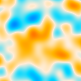} & \includegraphics[width=0.05\textwidth]{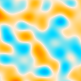} & \includegraphics[width=0.05\textwidth]{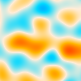} & \includegraphics[width=0.05\textwidth]{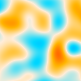} \\
\hline
\end{tabular}
\caption{10$\arcmin$x10$\arcmin$ 3-day thumbnail maps for each transient. The upper row is the intensity map with $\pm$5000~$\mu$K color range. The bottom row is the $S/N$ map after applying a matched filter, with $\pm$5 color range. Due to the conjugate gradient iteration used to solve the maximum-likelihood map-making equation only being run for 10 steps, these maps are effectively mildly highpass filtered. The affected scales have negligible weight in the matched filter. Events 2, 9, 10, 13 at the bottom of the table are the four events that are difficult to determine if they are real transients. }
\label{fig:3day}
\end{figure*}

For the rest of the candidates, we made $2^{\circ}$ by $2^{\circ}$ intensity maps (shown in Figure~\ref{fig:3day}) from TODs covering a three day period (or shorter if the scanning cadence is especially high) centered on each event. These maps, referred as ``thumbnail maps," have 0.25-arcminute resolution and are centered at the candidate positions. We visually inspected these maps both before and after matched filtering\footnote{This is pure map-space matched filtering, which should be more accurate than the mixed TOD-level/map-space matched filtering used in the main 3-day maps used for the initial search, but still depend on an accurate noise model.} and remove candidates with with an extended or irregular shape, since we expect all our transients to be point sources. This issue may be caused by poor data quality causing arcminute scale fluctuations. For a few of the remaining candidates, we observed that the raw and filtered maps appear discrepant with each other, e.g. a supposedly strong detection having no visible counterpart in the maps before filtering; or there was an overdensity of $> 3\sigma$ peaks near the event in the filtered maps, which should not happen if the noise model is correct. We do not discard these events, but classify them as low quality candidates. All in all, we are left with 29 transient detections. Eight of these events are due to asteroids (see section \ref{sub:characterization}), three were previously detected, and four are low quality candidates.


\section{Results} \label{sec:results}
\subsection{Characterization}
\label{sub:characterization}

\begin{figure*}[htbp]
     \includegraphics[width=0.44\linewidth]{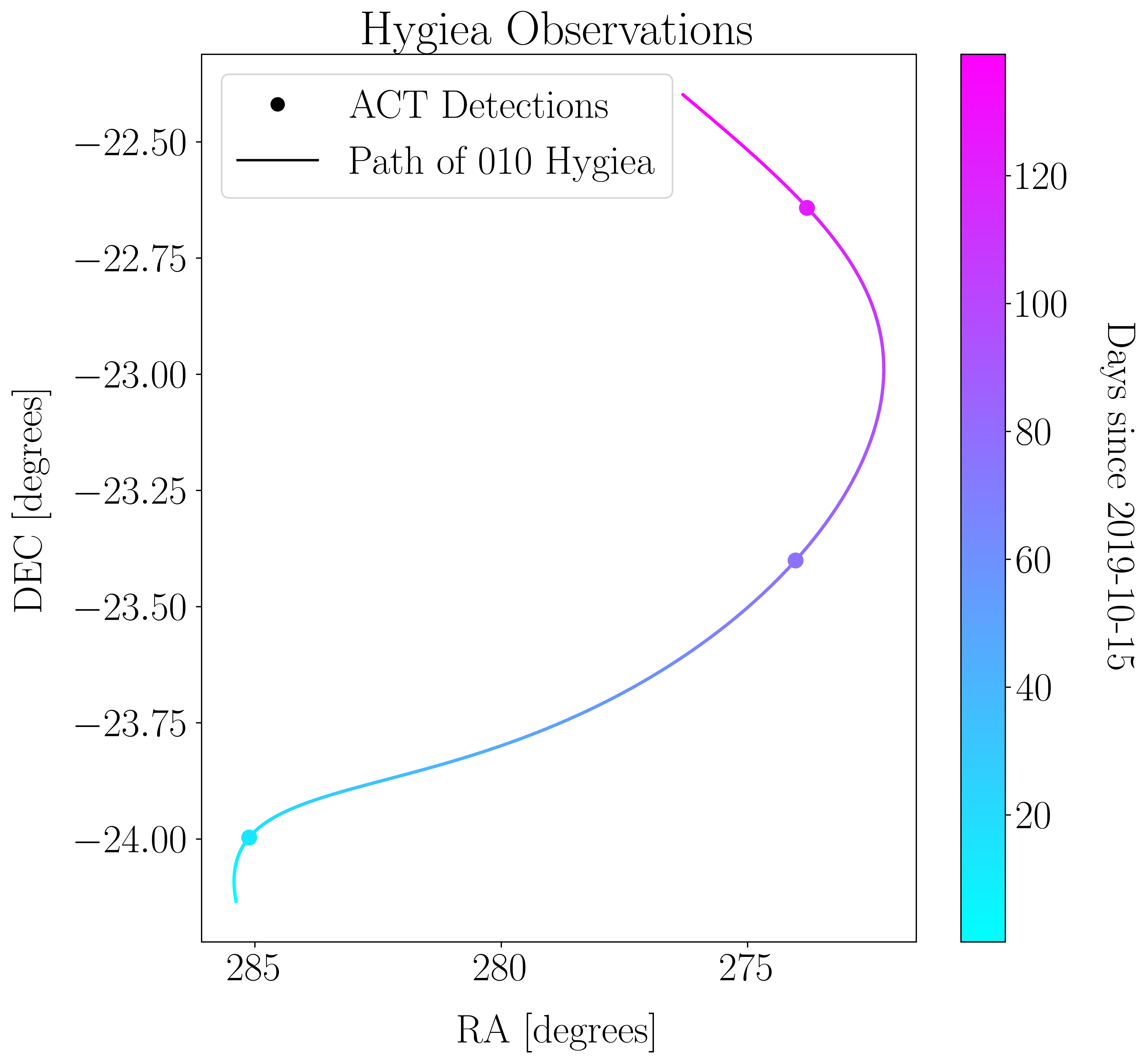}
     \includegraphics[width=0.44\linewidth]{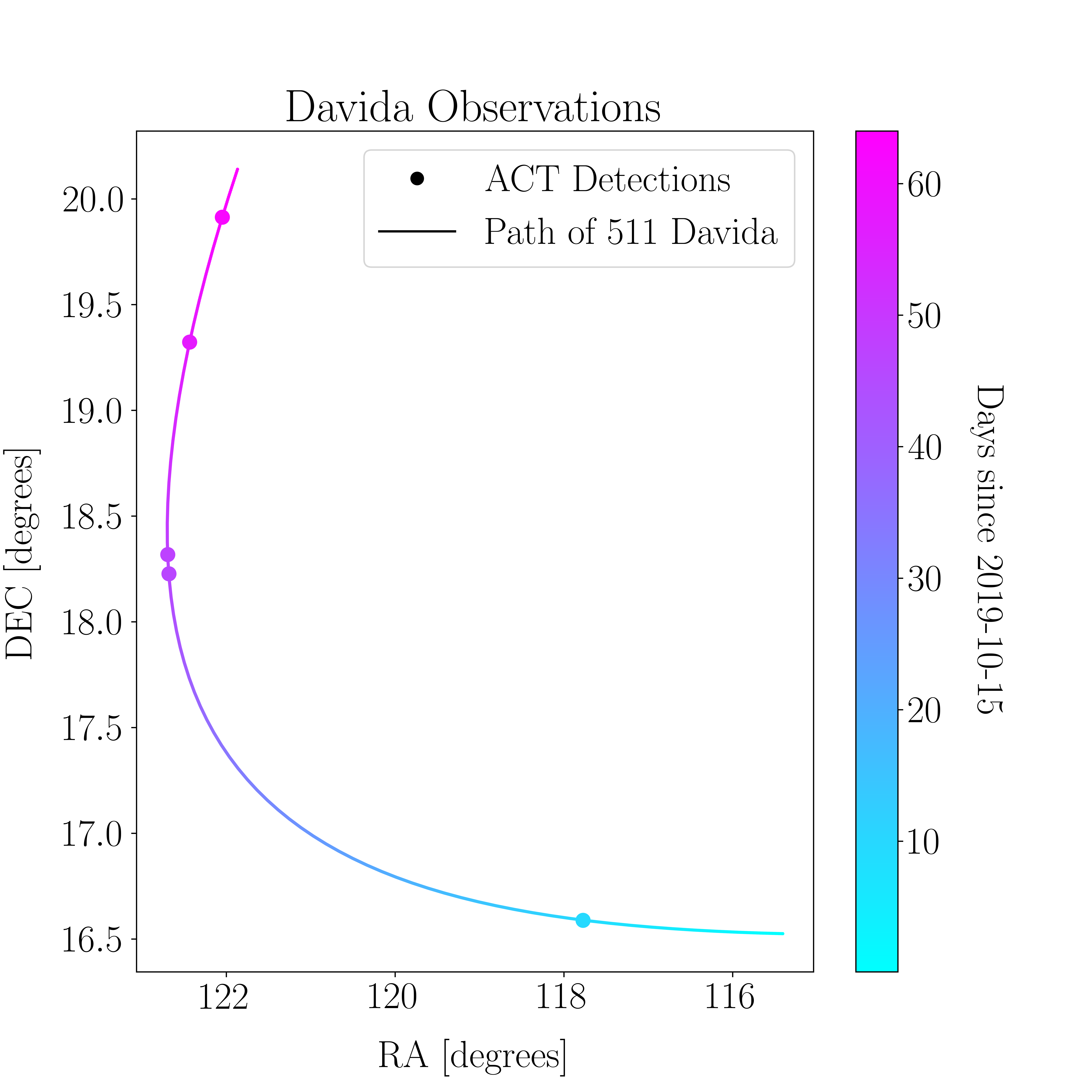}
    \caption{{\bf \textit{Left:}} The three transient events consistent with Hygiea observations. {\bf \textit{Right:}} The five transient events consistent with Davida observations. All of the plotted positions and observation times are consistent with the asteroids' paths. The position errors are on the order of 0.1 arcminutes.}
    \label{fig:asteroids}
\end{figure*}

Two groups of our transient events appeared to be close in time and space to each other (see Table ~\ref{tab:asteroid}). Using the IAU Minor Planet Center NEOChecker webtool,\footnote{See https://minorplanetcenter.net/cgi-bin/checkneo.cgi} we found three events to be coincident with asteroid 10 Hygiea and five events to be coincident with asteroid 511 Davida. In Figure~\ref{fig:asteroids}, we plot the paths of these asteroids with our events overlayed. All eight events appear consistent in space and time with the asteroid observations which were not masked in our maps. For future blind transient searches we must be prepared for the possibility of detections from moving objects such as asteroids. This will become especially important for real-time follow-up observations. A paper dedicated to studying asteroids in ACT data is in preparation.\footnote{Orlowski-Scherer et al., in prep.}

\begin{table*}[htb!]
    \centering
    \begin{tabular}{|l|l|l|l|l|l|l|l|l|l|l|l|}
    \hline
     \multirow{2}{*}{Name}& \multirow{2}{*}{RA $\degr$}   &  
     \multirow{2}{*}{Dec $\degr$} & Pos. &  
     \multicolumn {3}{c|}{Peak Flux (mJy)} &
     \multicolumn {3}{c|}{Mean Flux (mJy)}  & 
     Peak & \multirow{2}{*}{$\alpha$} \\
    \cline{5-11}
      & & &Acc.$\arcsec$&f220 & f150 & f090 & f220 & f150 &f090& Time (UTC) & \\
    \hline
    \hline
    \multirow{2}{*}{Hygiea(a)} & \multirow{2}{*}{285.1166} &  \multirow{2}{*}{$-$23.9962} &\multirow{2}{*}{14} & 334& 199 & 104&  1.4&1.3 &0.8 &2017-05-21  &1.4  \\
     &&&&$\pm$71 & $\pm$21& $\pm$20  & $\pm$4.0 &$\pm$1.5 &$\pm$2.6 &04:27 &  $\pm$0.3        \\
     \hline
    \multirow{2}{*}{Hygiea(b)} &  \multirow{2}{*}{274.0251} & \multirow{2}{*}{$-$23.4008}& \multirow{2}{*}{5}& 369&218 &92 & 19.1 &3.0 &0.5 &2017-07-24   & 1.5 \\
    &&&&$\pm$42 & $\pm$18& $\pm$20  &$\pm$6.7  &$\pm$2.6 &$\pm$4.6 &05:50& $\pm$0.2          \\
    \hline
    \multirow{2}{*}{Hygiea(c)} &\multirow{2}{*}{273.7942} & \multirow{2}{*}{$-$22.6426} &\multirow{2}{*}{10}&351&144&65& 18.5&8.5&$-$0.2&2017-09-09 &2.1                    \\
    &&&&$\pm$52 & $\pm$20& $\pm$19  & $\pm$7.3 &$\pm$2.8 &$\pm$5.0 &02:54 &$\pm$0.4          \\
    \hline
    \hline
    \multirow{2}{*}{Davida(a)}&\multirow{2}{*}{117.7733} &  \multirow{2}{*}{16.5879}&\multirow{2}{*}{8}&206&114&53&4.0&1.9&$-$1.9&2019-10-24&1.6   \\  
    &&&&$\pm$34 & $\pm$15& $\pm$14  & $\pm$2.1 &$\pm$0.8 &$\pm$1.4 & 09:01&$\pm$0.3 \\
    \hline
    \multirow{2}{*}{Davida(b)} &\multirow{2}{*}{122.6793} &  \multirow{2}{*}{18.2278}&\multirow{2}{*}{5}& 277&155&57&$-$0.8&0.2&4.0&2019-11-30 &1.8\\ 
    &&&&$\pm$32 & $\pm$13& $\pm$12  & $\pm$ 2.1&$\pm$0.8 &$\pm$1.4 &07:12 &  $\pm$0.2 \\
    \hline
    \multirow{2}{*}{Davida(c)}    &  \multirow{2}{*}{122.6946} &  \multirow{2}{*}{18.3167}&\multirow{2}{*}{8}&192 &118&57&$-$2.2&0.3&3.1&2019-12-01 &1.6                 \\
    &&&&$\pm$39 & $\pm$16& $\pm$13  & $\pm$2.0 &$\pm$0.8 &$\pm$1.3 & 08:57&  $\pm$0.4 \\
    \hline
    \multirow{2}{*}{Davida(d)} &  \multirow{2}{*}{122.4329} &  \multirow{2}{*}{19.3231}&\multirow{2}{*}{2}&343 &162&86&5.1&1.1&3.3&2019-12-11 &1.7                     \\
    &&&&$\pm$27 & $\pm$14& $\pm$10  & $\pm$2.0 &$\pm$0.7 &$\pm$1.3&06:45 & $\pm$0.2 \\
    \hline
    \multirow{2}{*}{Davida(e)}&  \multirow{2}{*}{122.0468} &  \multirow{2}{*}{19.9151}&\multirow{2}{*}{6}&313 &174&62&$-$5.1&0.8&1.8&2019-12-16&1.8\\
    &  &  &&$\pm$44 & $\pm$16& $\pm$15  &$\pm$1.9  &$\pm$0.8 &$\pm$1.4 &05:10 & $\pm$0.3 \\
    \hline
    \end{tabular}
    \caption{Eight events associated with asteroids.}
    \label{tab:asteroid}
\end{table*}

The properties of the remaining 21 transient events are shown in Table~\ref{tab:characterization}. Three single detections labeled as ``N"  were previously published in~\cite{Naess_2021a}. Among the newly detected events, 12 are single events and three repeat twice, with time intervals ranging from one month to one year. The low (1 arcminute) resolution of the 3-day maps potentially increases the candidates’ position error, so we re-estimate the candidate positions using a method similar to the one used in the initial detection on the higher-resolution thumbnail maps described in Section~\ref{subsec:verfication}. We first subtract the corresponding f220/f150/f090 mean sky map and apply a matched filter. The final position for each candidate is the inverse-variance-weighted average among individual positions, using the ratio between half maximum of the beam and the appropriate $S/N$ as weights. The position accuracy is evaluated by the square root of the variance of weighted mean. As mentioned in Section~\ref{subsec:verfication}, four out of the 12 single events are difficult to determine if they are real transients due to unusual noise patterns on the intensity maps. We present these events at the bottom of the table without assigning a formal name.
\begin{table*}[htb!]
\footnotesize
\centering
    \begin{tabular}{|l|l|l|l|l|l|l|l|l|l|l|l|l|l|l|}
    \hline
    \multirow{2}{*}{Ind}&Name& \multirow{2}{*}{RA $^{\circ}$}   &  
     \multirow{2}{*}{Dec $^{\circ}$} & Pos. &  
     \multicolumn{3}{c|}{Peak Flux (mJy)} &
     \multicolumn{3}{c|}{Mean Flux (mJy)}  & 
     \multicolumn{3}{c|}{Time} & \multirow{2}{*}{$\alpha$} \\
    \cline{6-14}
      &(ACT-T)& & &Acc.$\arcsec$&f220 &f150 & f090 & f220&f150&f090& Peak (UTC) &Rise&Fall& \\
    \hline
    \hline   
    1& J060702&\multirow{2}{*}{91.7603} &\multirow{2}{*}{17.6993} &\multirow{2}{*}{22}&36&78&64&$-1.2$&$-2.3$&$-4.6$&2017-09-05 &\multirow{2}{*}{$<43\,$min}&\multirow{2}{*}{$<25\,$h}&0.0\\
    &+174157&&&&$\pm$35 & $\pm$13& $\pm$14  &$\pm$2.3  &$\pm$0.9 &$\pm$1.6 &05:35 & & & $\pm$0.5\\
    \hline
    3& J224500&\multirow{2}{*}{341.2510} &\multirow{2}{*}{$-$33.2589} &\multirow{2}{*}{14}&192&253&255&5.5&4.4&8.8&2017-10-08 &\multirow{2}{*}{$<1\,$d}&\multirow{2}{*}{$>23\,$min}&$-$0.6\\
    &-331532&&&&$\pm$41 & $\pm$20& $\pm$12  &$\pm$2.3  &$\pm$0.8 &$\pm$1.4 &18:17 & & & $\pm$0.2\\
    \hline
    \multirow{2}{*}{4a}&&\multirow{4}{*}{343.2597} &\multirow{4}{*}{16.8408} &\multirow{4}{*}{8}&1036&1154&801&&&&2018-09-10 &\multirow{2}{*}{}&\multirow{2}{*}{$<2\,$d}&0.5\\
    &J225302&&&&$\pm$122 & $\pm$52& $\pm$34  &4.7  &7.8 &18.7 &21:55 & & & $\pm$0.1\\
    4b&+165027& & &&$-$15&141&148&$\pm$3.2&$\pm$1.1&$\pm$2.0&2019-06-06 &\multirow{2}{*}{$<2\,$d}&\multirow{2}{*}{$<3\,$d}&$-$0.6\\
    &&&&&$\pm$58 & $\pm$20& $\pm$18  &  & & &04:18 & & & $\pm$0.4\\
    \hline
    \multirow{2}{*}{5a}& &\multirow{4}{*}{292.1328} &\multirow{4}{*}{$-$35.1327}&\multirow{4}{*}{3}&2496&2383&1525&&&&2018-10-04 &\multirow{2}{*}{$<1\,$d}&\multirow{2}{*}{$<20\,$d}&0.9 \\
    &J192831&&&&$\pm$177 & $\pm$40& $\pm$27  & 4.7&11.7&37.2 &02:22 & & & $\pm$0.05         \\
    5b&-350757   &   & &&545 &645&490&$\pm$3.7 &$\pm$1.4 &$\pm$2.7 &2019-08-10&\multirow{2}{*}{$<3\,$d}&\multirow{2}{*}{$>1\,$d}&0.4 \\
    &&&&&$\pm$69 & $\pm$23& $\pm$20  &  & & &23:10 & & & $\pm$0.1          \\
    \hline
    6&J190222&\multirow{2}{*}{285.5938} &\multirow{2}{*}{$-$5.6028} &\multirow{2}{*}{10}&926&717&350&12.7&5.3&7.7&2018-10-20 &\multirow{2}{*}{$<1\,$d}&\multirow{2}{*}{$<1\,$d}&1.2\\
    &-53610&&&&$\pm$87 & $\pm$34& $\pm$29  &$\pm$5.6  &$\pm$2.1 &$\pm$3.9 &20:32 & & & $\pm$0.1\\
    \hline
    7& J085813&\multirow{2}{*}{134.5579} &\multirow{2}{*}{19.7630} &\multirow{2}{*}{4}&103&197&269&3.0&0.1&2.4&2018-11-15 &\multirow{2}{*}{$<5\,$d}&\multirow{2}{*}{$<22\,$h}&$-$0.8\\
    &+194546&&&&$\pm$31 & $\pm$9& $\pm$9  &$\pm$1.9  &$\pm$0.8 &$\pm$1.5 &05:18 & & & $\pm$0.1\\
    \hline
    8&J142555&\multirow{2}{*}{216.4831} &\multirow{2}{*}{14.2020} &\multirow{2}{*}{8}&446&563&562&$-$0.5&1.7&5.6&2018-11-21 &\multirow{2}{*}{$<20\,$h}&\multirow{2}{*}{$>23\,$h}&$-$0.1\\
    &+141207&&&&$\pm$61 & $\pm$17& $\pm$14  &$\pm$1.9  &$\pm$0.6 &$\pm$1.1 &08:07 & & & $\pm$0.1\\
    \hline
    11&J060757&\multirow{2}{*}{91.9890} &\multirow{2}{*}{$-$54.4408} &\multirow{2}{*}{20}&108&113&113&-3.6&0.8&2.6&2019-08-09 &\multirow{2}{*}{$<24\,$h}&\multirow{2}{*}{$<18\,$h}&0.0\\
    &-542626&&&&$\pm$65 & $\pm$21& $\pm$15  &$\pm$2.6  &$\pm$1.0 &$\pm$2.1 &12:18 & & & $\pm$1.6\\
    \hline
    \multirow{2}{*}{12a}& &  \multirow{4}{*}{54.1961} &\multirow{4}{*}{0.5865}&\multirow{4}{*}{5}&108&302&448&&&&2019-08-10&\multirow{2}{*}{$<30\,$h}&\multirow{2}{*}{$>23\,$h}& $-$1.1                   \\
    &J033647&&&&$\pm$66 & $\pm$26& $\pm$25  &11.7&11.9&30.1 &13:35 & & & $\pm$0.2 \\
    12b&+03511 &&&&356 &487&580 &$\pm$4.0  &$\pm$1.5 &$\pm$2.8&2019-09-28&\multirow{2}{*}{$>2\,$d}&\multirow{2}{*}{$<2\,$d}&  $-$0.5                  \\
    &&&&&$\pm$67 & $\pm$23& $\pm$21  &  & & &10:23 & & & $\pm$0.1 \\
    \hline
    14&J125045&\multirow{2}{*}{192.6887} &\multirow{2}{*}{11.5607} &\multirow{2}{*}{14}&76&202&218&$-$1.8&1.8&5.4&2019-09-17 &\multirow{2}{*}{$<1\,$d}&\multirow{2}{*}{$<49\,$d}&$-$0.3\\
    &+113338&&&&$\pm$114 & $\pm$29& $\pm$21  &$\pm$1.8  &$\pm$0.6 &$\pm$1.2 &11:34 & & & $\pm$0.4\\
    \hline
    15&J180723&\multirow{2}{*}{271.8483} &\multirow{2}{*}{19.7063} &\multirow{2}{*}{14}&\multirow{2}{*}{NA}&82&153&0.5&2.8&5.5&2019-11-13 &\multirow{2}{*}{$<6\,$d}&\multirow{2}{*}{$<12\,$d}&$-$1.4\\
    &+194222&&&& & $\pm$24& $\pm$19  &$\pm$2.2  &$\pm$0.8 &$\pm$1.6 &13:01 & & & $\pm$0.7\\
    \hline
    \hline
    N1&J181515&\multirow{2}{*}{273.8166} &\multirow{2}{*}{$-$49.4627} &\multirow{2}{*}{5}&\multirow{2}{*}{NA}&555&282&$-$2.5&6.4&16.0&2019-11-08 &\multirow{2}{*}{$>8\,$min}&\multirow{2}{*}{$>4\,$min}&1.3\\
    &-492746&&&& & $\pm$28& $\pm$18  &$\pm$5.7  &$\pm$2.2 &$\pm$4.7 &17:22 & & & $\pm$0.2\\
    \hline
    N2&J070038&\multirow{2}{*}{105.1588} &\multirow{2}{*}{$-$11.2458} &\multirow{2}{*}{10}&\multirow{2}{*}{NA}&344&152&$-$1.8&5.5&4.5&2019-12-14 &\multirow{2}{*}{$<8\,$d}&\multirow{2}{*}{$>8\,$min}&1.9\\
    &-111436&&&& & $\pm$30& $\pm$21  &$\pm$7.2  &$\pm$2.7 &$\pm$5.2 &21:30 & & & $\pm$0.5\\
    \hline
    N3&J200758&\multirow{2}{*}{301.9965} &\multirow{2}{*}{16.1642} &\multirow{2}{*}{7}&222&300&346&-2.2&2.7&15.4&2018-09-11 &\multirow{2}{*}{$<1\,$d}&\multirow{2}{*}{$>3\,$d}&-0.4\\
    &+160954&&&&$\pm$53 & $\pm$26& $\pm$18  &$\pm$3.0  &$\pm$1.1 &$\pm$2.1 &19:36 & & & $\pm$0.2\\
    \hline
    \hline 
    2& &\multirow{2}{*}{316.0950} &\multirow{2}{*}{$-$13.6699} &\multirow{2}{*}{20}&\multirow{2}{*}{NA}&1050&1311&0.9&$-$2.7&$-$2.0&2017-09-27 &\multirow{2}{*}{$<6\,$min}&\multirow{2}{*}{$<1\,$d}&$-$0.5\\
    &&&&& & $\pm$56& $\pm$30  &$\pm$4.5  &$\pm$1.6 &$\pm$3.0 &00:21 & & & $\pm$0.1\\
    \hline
    9&&\multirow{2}{*}{196.7753} &\multirow{2}{*}{16.6089} &\multirow{2}{*}{21}&168&117&12&0.6&0.0&1.2&2019-07-19 &\multirow{2}{*}{$<3\,$h}&\multirow{2}{*}{$>22\,$h}&2.1\\
    &&&&&$\pm$52 & $\pm$19& $\pm$16  &$\pm$1.7  &$\pm$0.6 &$\pm$1.1 &19:23 & & & $\pm$0.6\\
    \hline
    10&&\multirow{2}{*}{208.4267} &\multirow{2}{*}{6.7745} &\multirow{2}{*}{23}&293&200&110&$-$1.5&0.2&0.3&2019-08-05 &\multirow{2}{*}{$<19\,$h}&\multirow{2}{*}{$<1\,$d}&1.2\\
    &&&&&$\pm$57 & $\pm$21& $\pm$19  &$\pm$2.0  &$\pm$0.7 &$\pm$1.3 &14:40 & & & $\pm$0.3\\
    \hline
    13&&\multirow{2}{*}{60.9932} &\multirow{2}{*}{14.0094} &\multirow{2}{*}{26}&920&542&603&$-$2.1&1.7&1.2&2019-08-14 &\multirow{2}{*}{$<5\,$min}&\multirow{2}{*}{$<2\,$min}&1.6\\
    &&&&&$\pm$103 & $\pm$52& $\pm$44  &$\pm$3.3  &$\pm$1.2 &$\pm$2.3 &09:27 & & & $\pm$0.2\\
    \hline 
    \end{tabular}
\caption{Properties of the transient events. The candidates are listed in the order of the detection time (or the detection of the first event for the repeating candidates). Three single detections labeled as ``N" are the redetections of the transients published in \cite{Naess_2021a}. It is difficult to fit for rise and fall time using light curves due to uneven scanning cadence. Instead, we examine if the scan right before and after the scans with the peak flux density has a $>5\sigma$ detection, and calculate the time interval in between the scans. The rise time of Event 4a is left empty because the time gap between the peak scan and the scan right before is longer than 50 days. Spectral index is evaluated as described in text, using array-wise flux density values, except for candidate 13. This transient event completed the rise and fall process within the time the sky took to drift across the array. We therefore evaluate the spectral index using flux values taken by only one quarter of detectors in each array, to capture the peak flux.}
\label{tab:characterization}
\end{table*}

\begin{figure*}[htbp]
    \centering
    \includegraphics[width=0.95\textwidth]{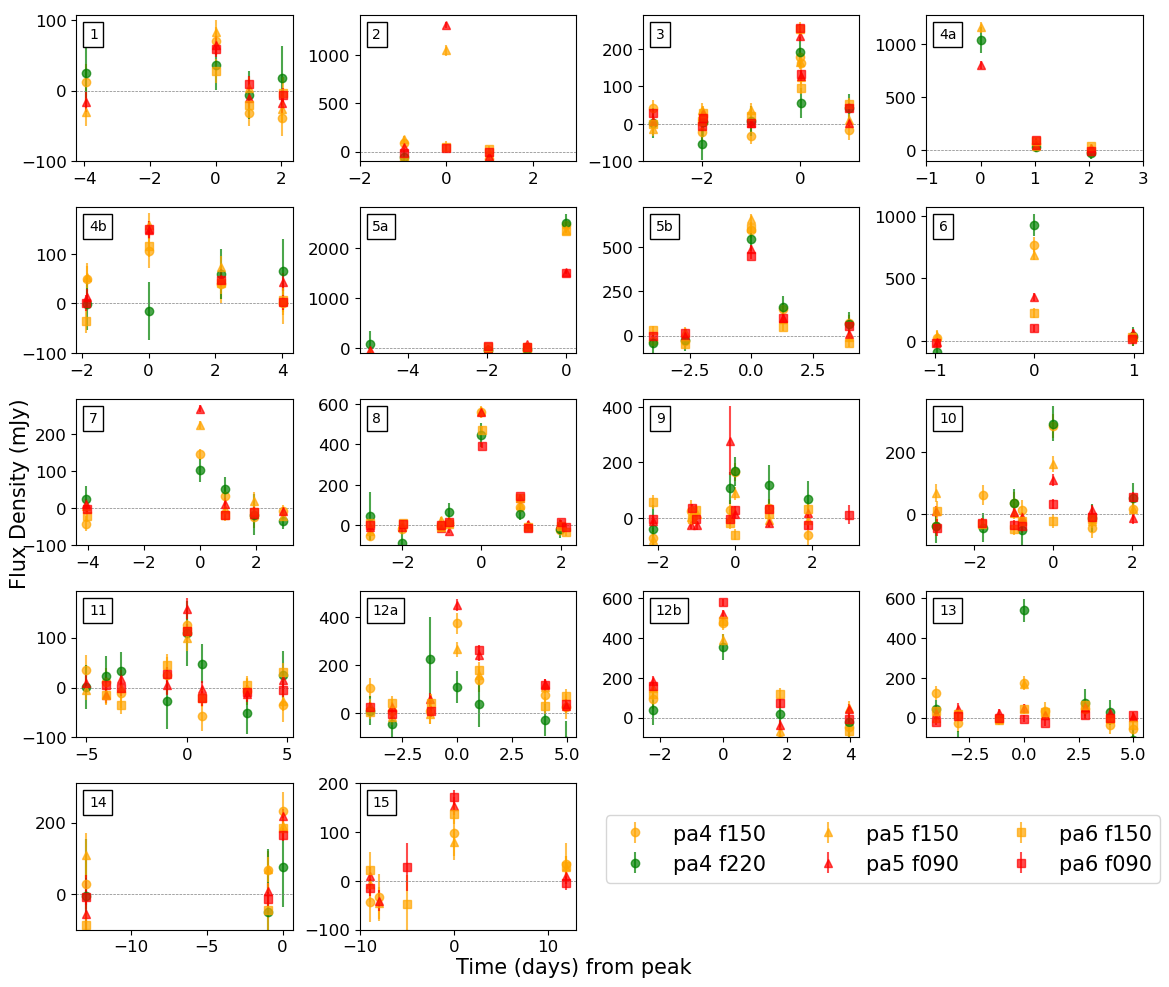}
    \caption{Light curves for 18 detected transient events on day time scales from the peak. Each frequency is denoted by a different color. We see that in most cases the peak is correlated with all frequencies.}
    \label{fig:short_light_curve}
\end{figure*}

\begin{figure*}[htbp]
    \centering
    \includegraphics[width=0.95\textwidth]{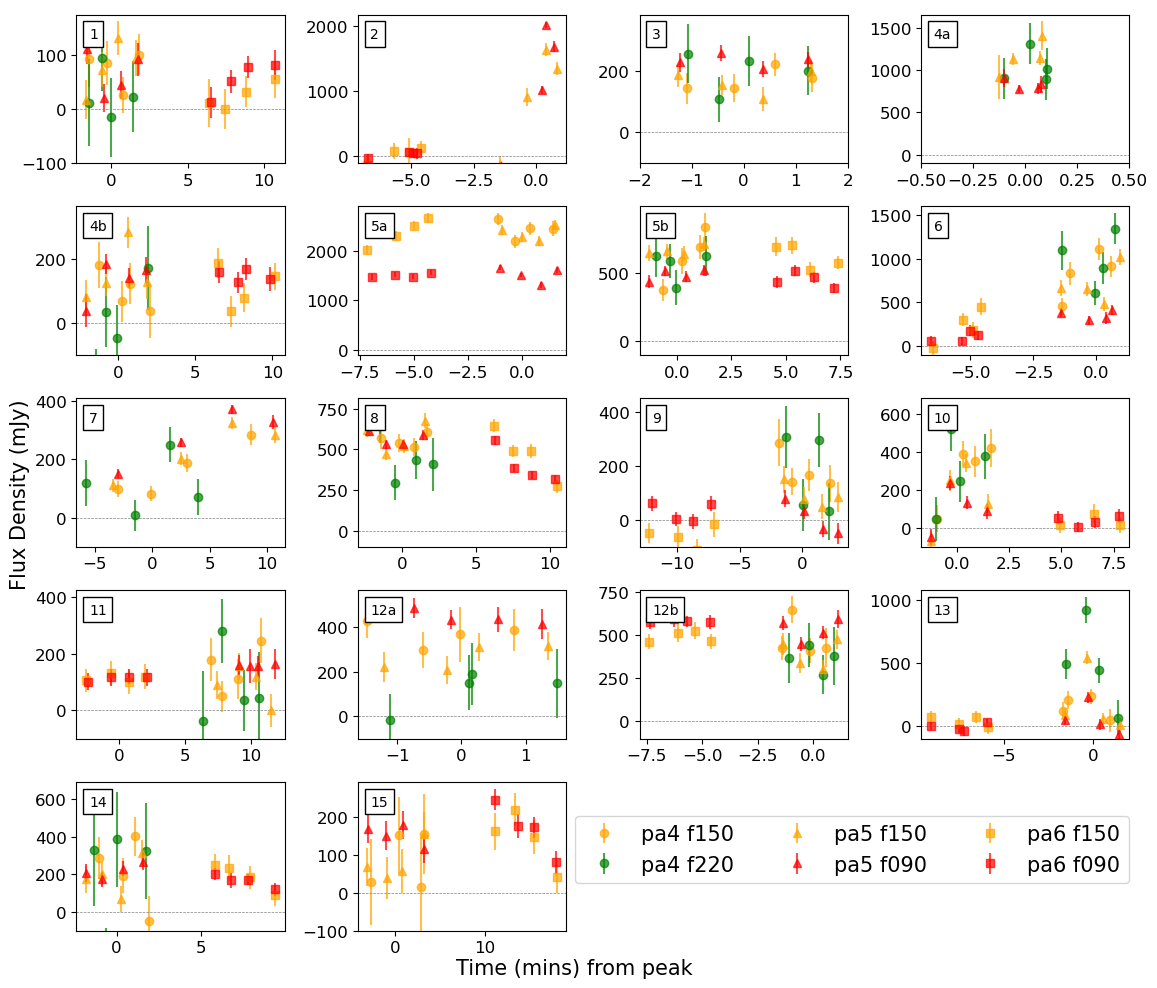}
    \caption{High-resolution light curves for transient events on minute time scales. Events 1, 2, 5, 6, 7, 8, 9, 10, 13, 14 and 15 show strong signal variation within a space of four minutes.}
    \label{fig:short_light_curve_subarray}
\end{figure*}

Flux densities for each scan are evaluated at the refined positions and used to generate light curves  (Figure~\ref{fig:short_light_curve}) using the same method described in Section~\ref{subsec:verfication}. Higher-resolution light curves (Figure~\ref{fig:short_light_curve_subarray}) are made by dividing detectors in each array into four groups by the order of observation time to study the minute by minute change in flux density. When the flux density values are available in both bands of PA4 and PA5, the spectral index $\alpha$ is evaluated by the best-fit of a power law $S_{\nu}\propto\nu^{\alpha}$ of frequency $\nu$ to the flux measurement $S_{\nu}$ at the three different bands, taking into account the variance of each band flux. Under this circumstance, PA6 is omitted because it scans across the candidates approximately 10 minutes earlier (or later when the sky is rising) than PA4 and PA5. Events 2 and 15 do not have f220 measurements, so $\alpha$ is directly calculated by 
\begin{equation}
    \alpha=\frac{\log(S_{\nu_{1}})-\log(S_{\nu_{2}})}{\log(\nu_{1})-\log(\nu_{2})}
\end{equation}
using the f150 and f090 flux measured by PA5 or PA6, depending on which array measures the peak flux.

\subsection{Counterparts}

We searched for counterparts for each transient event in the SIMBAD\footnote{https://simbad.cds.unistra.fr/simbad/} \citep{SIMBAD} database. The results of this search are summarized in Table~\ref{tab:counterparts}. We searched for any Gaia objects \citep{Gaia_dr3} within one arcminute of each event and list the SIMBAD identification. We calculate the probability of a chance association given the density of Gaia stars \citep{Gaia_dr3} with the counterpart's magnitude or lower within the counterpart's separation of the transient's position. We exclude any counterparts with a chance association probability greater than ten percent. We also searched for counterparts of the transients found in \cite{Naess_2021a} that are recovered in this analysis. The associations agree with the original findings.

In addition to the Gaia source catalog, we search for galactic objects in the Gaia extragalactic catalog \citep{Gaia_dr3} using the VizieR \citep{Ochsenbein_2000}. Only events 10 and 14 have associations from this catalog. Event 10 is coincident with an AGN with a 1.1 percent chance of a false association (see Table~\ref{tab:counterparts}). Event 14 is coincident with a RR Lyrae star with a 2.4 percent chance of a false association. This association is not listed because there is a much more obvious Gaia star associated with event 14.

\begin{figure*}
    \centering
    \includegraphics[width=\textwidth]{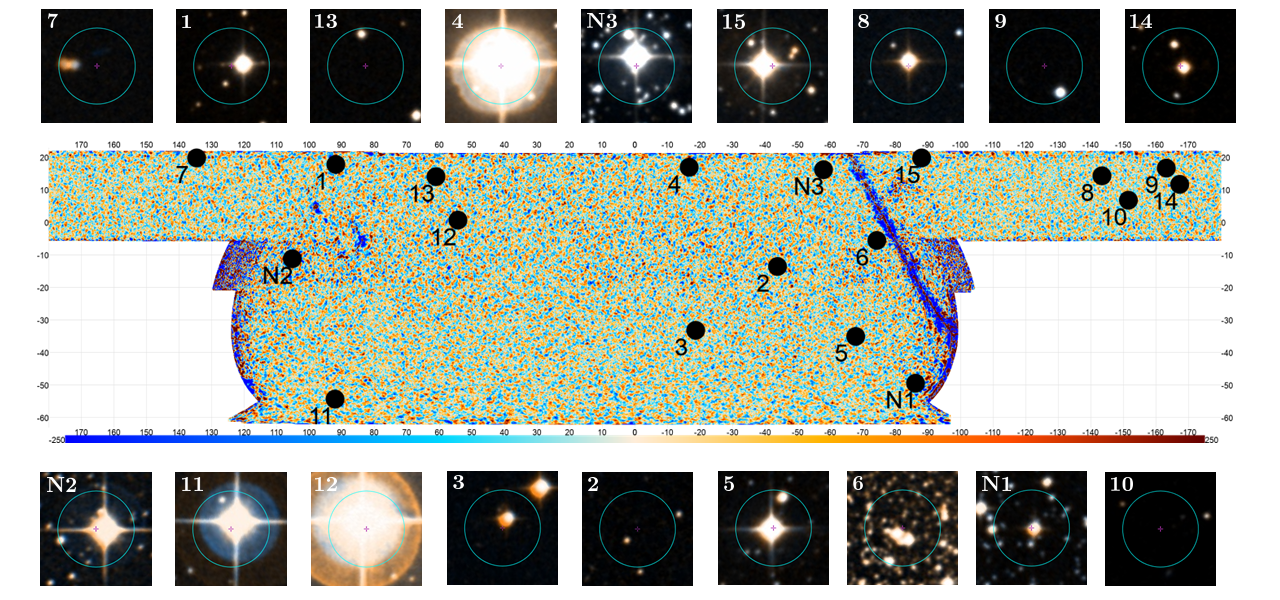}
    \caption{Combined ACT data from 2007--2019 overlayed with the transient positions. Images from DSS at each location, indicated by a in the center of each image, are also shown with a one arcminute contour plotted for scale. The majority of the transient events have bright stellar counterparts seen in the optical.}
    \label{fig:coadd_dss}
\end{figure*}

In Figure~\ref{fig:coadd_dss} we plot the positions of the transients overlayed on combined ACT data from 2007--2019. This plot includes color images in the optical from the Digitized Sky Surveys (DSS)\footnote{https://archive.stsci.edu/cgi-bin/dss\_form}\footnote{These images were acquired using the online tool Aladin Lite. See \cite{Aladin}.} that show there are bright stellar counterparts for most of the events.

\begin{table*}[ht!]
\centering
\begin{tabular}{|l|l|l|l|l|l|l|}
\hline
Name & ID & Object Type & Magnitude & Pos Err (") & Sep (") & Chance \\
\hline
\hline
1 & V* CP Ori & Eclipsing Binary, G0 & 10.52 & 22 & 8.60 & 5.66e-04  \\
\hline
3 & V* TX PsA & Eruptive Variable, M5IVe & 11.84 & 14 & 4.24 & 1.21e-04  \\
\hline
4a/b & V* IM Peg & RS CVn Variable, K2III & 5.66 & 8 & 1.48 & 4.23e-08  \\
\hline
5a/b & HD 182928 &  Rotating Variable, G8IIIe & 9.37 & 3 & 2.15 & 5.79e-06  \\
\hline
6 & ** SKF 1810A & Young Stellar Object Candidate & 12.73 & 10 & 7.00 & 5.05e-03  \\
\hline
7 & G 9-38 & High Proper Motion Binary, M7V/M8Ve & 12.49/11.97 & 4 & 4.03 & 1.56e-04  \\
\hline
8 & StKM 1-1155 & Low-mass Star, M0.0Ve & 10.91 & 8 & 2.71 & 1.78e-05  \\
\hline
11 & V* TY Pic & RS CVn Variable, G8/K0III+F & 7.29 & 20 & 6.80 & 9.82e-06  \\
\hline
12a/b & HD 22468 & RS CVn Variable, K2:Vnk/K4 & 5.60/8.51 & 5 & 4.10 & 2.59e-07  \\
\hline
14 & Gaia DR2 3927810990205301504  & Star & 12.43 & 14 & 2.88 & 5.96e-05  \\
\hline
15 & HD 347929 & Rotating Variable, K0 & 9.04 & 14 & 7.66 & 8.04e-05  \\
\hline
\hline
N1 & 2MASS J18151564-4927472 & High Proper Motion Star, M3 & 11.72 & 5 & 4.25 & 5.47e-04  \\
\hline
N2 & HD 52385 & Star, K0/1III & 8.11 & 10 & 7.46 & 5.15e-05  \\
\hline
N3 & HD 191179 & Spectroscopic Binary, G5 & 7.96 & 7 & 7.05 & 4.41e-05  \\
\hline
\hline
2 & Gaia DR2 6885400713762009216 & Star & 16.74 & 20 & 13.02 & 5.96e-02  \\
\hline
9 & Gaia DR2 3936693910286240000 & Star & 13.87 & 21 & 33.51 & 2.11e-02  \\
\hline
10\footnote{Field too dense to identify stellar counterpart} & 1636148068921376768 & AGN & -- & 23 & 17.40 & 1.20e-02  \\
\hline
13 & Gaia DR2 38908044312695424 & Star & 14.34 & 26 & 23.58 & 2.51e-02  \\
\hline

\end{tabular}
\caption{Possible counterparts for each transient event from the SIMBAD database. If known, the spectral type is given next to the object type. The chance of a false association is calculated using the density of Gaia sources with the same magnitude or brighter of the counterpart. If the Gaia object is not found in the SIMBAD database, then the Gaia identification number is listed. All separations are calculated using the Gaia coordinates. This table also includes counterparts associated with the three previously published ACT transients from \cite{Naess_2021a}. When two stars from the same system are resolved, as is the case with events 7 and 12, we quote the average separation weighted by one minus the chance association and we list the lowest of the two chance associations.}
\label{tab:counterparts}
\end{table*}

\section{Discussion and Conclusions} \label{sec:concl}
We have demonstrated a pipeline of transient detection and characterization using ACT data, and presented a catalog of 21 millimeter transients from 3-day maps, including three of which that were previously detected, and four events that cannot be confirmed as real transients. After assessing the spectral indices and counterpart associations of these events, two classes of objects emerge: flares associated with stars with flat or falling spectra, indicating radiation from sychrotron emission, and stellar flares with rising spectra, indicating thermal emission. Performing a statistical analysis on such a small sample size is difficult, but in the future as we discover more millimeter stellar transients, we will be able to better characterize them. For example, we will be able to determine whether these events reside near the galactic plane or scattered across the sky, or whether two classes of transients, rising versus falling spectra, really exist or if there is simply a wide distribution of spectral indices. Tests of these statistics with our small number of events are inconclusive. 

Events 3, 4a, 4b, 5a, 5b, 11, 12a, 12b, and 15 are associated with rotating variable stars. These stars have large dark spots on their surface from magnetic activity in their chromospheres. These spots cause intrinsic variability in flux in optical and IR wavelengths but also cause flares in radio wavelengths from synchrotron radiation \citep{hjellming_1980}. Events 4, 11, 12, and 15 are coincident with a special type of rotating variable stars called RS Canum Venaticorum variables (RSCVn). These are binary systems that also exhibit variability due to dark spots (See \citealt{hall_1976}, \citealt{Zeilik_1979}). Similar events were seen in the millimeter systematic transient search from SPT-3G \citep{Guns_2021}. 

Event 1 is also associated with a binary system. Although it is not classified as rotationally variable, it most likely flares in the microwave with a similar mechanism. This association is similar to ACT-T J200758+160954 (N3) which was previously published in \cite{Naess_2021a}. 

Events 7, 8, and 15 are all coincident with cool G, K, and M type stars. Event 14 is associated with a Gaia star but its spectral type is unknown. Since all of these events also have negative spectral indices, these flares are most likely from magnetic activity producing synchrotron radiation. This is consistent with findings from \cite{Guns_2021} who reported microwave flares from M and K dwarfs. These cool stars have convective envelopes which cause increased magnetic activity but do not exhibit the same dark spots found in rotationally variable stars \citep{yang_2017}. 

Events 6, N1, and N2 all have rising spectra which are associated with thermal emission. In these cases, some mechanism other than synchrotron radiation is driving the flares. N1 and N2 are associated with cool M and K stars respectively but the spectral types of the star associations for 6 and 9 are unknown. \cite{Guns_2021} also observed two events with M-dwarf counterparts with rising spectra indicating this is a common class of millimeter transients. 

Although we cannot be certain events 2, 9, 10, and 13 are real transients, we still present possible counterparts for them in Table~\ref{tab:counterparts}. There is a six percent chance that the Gaia counterpart for Event 2 is a chance association. If the association is correct this transient is likely due to synchrotron radiation from stellar magnetic fields. The field around Event 10 is too dense to pinpoint a stellar counterpart. However, there is a small chance this event is associated with an AGN. \cite{Eftekhari_2022} predicts ACT will see on the order of ten gamma ray bursts but this event is not associated with any known GRB flares listed in the Fermi All-sky Variability Analysis \citep{Abdollahi_2017}. The associations for Events 9 and 13, are also unclear. Both events have rising spectra so they would most likely be similar to events 6, N1, and N2. The Zwicky Transient Facility (ZTF; \citealt{Masci_2018}) detected a flare with the same Gaia counterpart associated with event 13 twelve days after and 23$\arcsec$ away from the ACT detection. ZTF also detects repeating flares that are  34$\arcsec$ away from Event 9, although they did not occur close in time to the ACT detection.\footnote{This cross detection was done using software developed by \cite{Matheson_2021}}.

This sample likely only represents a fraction of microwave transient events present in the ACT data. The 3-day maps were not made for a systematic transient search. Recently, single observation maps that span more seasons than the 3-day maps have been made for the purpose of ACT time domain study. Major advantages of the single observation maps include the relative consistency in noise performance in each map and the freedom of stacking single scan maps for different periods of time. We will perform a similar analysis to search for transient events within those maps to gain a more accurate event rate for ACT. This will inform current and future large sky-area millimeter surveys such as The South Pole Telescope \cite{Carlstrom_2011}, Simons Observatory \citep{Ade_2019}, CCAT-prime \citep{Prime} and CMB-S4 \citep{Abazajian_2019} on how to best detect and study these events.

\section{Acknowledgements}
The Authors would like to thank Carles Badenes for his helpful discussions and comments.

This work was supported by the U.S. National Science Foundation through awards AST-0408698, AST-0965625, and AST-1440226 for the ACT project, as well as awards PHY-0355328, PHY-0855887 and PHY-1214379. Funding was also provided by Princeton University, the University of Pennsylvania, and a Canada Foundation for Innovation (CFI) award to UBC. ACT operates in the Parque Astron\'{o}mico Atacama in northern Chile under the auspices of the Agencia Nacional de Investigaci\'{o}n y Desarrollo (ANID; formerly Comisi\'{o}n Nacional de Investigaci\'{o}n Cient\'{i}fica y Tecnol\'{o}gica de Chile, or CONICYT).

Computations were performed using Princeton Research Computing resources at Princeton University. This research has made use of data and/or services provided by the International Astronomical Union's Minor Planet Center. This research has made use of the SIMBAD database, operated at CDS, Strasbourg, France.

YL acknowledges support from KIC Postdoctoral Fellowship. This work was supported by a grant from the Simons Foundation (CCA 918271, PBL). EC acknowledges support from the European Research Council (ERC) under the European Union’s Horizon 2020 research and innovation programme (Grant agreement No. 849169). SKC acknowledges support from NSF award AST-2001866. R.D. thanks CONICYT for grant BASAL CATA FB210003. CHC acknowledges FONDECYT Postdoc fellowship 3220255. ADH acknowledges support from the Sutton Family Chair in Science, Christianity and Cultures and from the Faculty of Arts and Science, University of Toronto. KMH acknowledges support by NSF awards AST-1815887 and AST-2206344. JPH gratefully acknowledges support from the estate of George A. and Margaret M. Downsbrough. CS acknowledges support from the Agencia Nacional de Investigaci\'on y Desarrollo (ANID) through FONDECYT grant no.\ 11191125 and BASAL project FB210003. EMV acknowledges support from NSF award AST-2202237.

The Digitized Sky Surveys were produced at the Space Telescope Science Institute under U.S. Government grant NAG W-2166. The images of these surveys are based on photographic data obtained using the Oschin Schmidt Telescope on Palomar Mountain and the UK Schmidt Telescope. The plates were processed into the present compressed digital form with the permission of these institutions. This research has made use of ``Aladin sky atlas'' developed at CDS, Strasbourg Observatory, France. This research has made use of the VizieR catalogue access tool, CDS, Strasbourg, France (DOI : 10.26093/cds/vizier). The original description of the VizieR service was published in 2000, AAS 143, 23

\clearpage
\bibliography{references}{}
\bibliographystyle{aasjournal}



\end{document}